\documentclass[aps,superscriptaddress,nofootinbib]{revtex4}%
\usepackage{graphicx}
\usepackage{amsmath}
\usepackage{amsfonts}
\usepackage{amssymb}%
\providecommand{\U}[1]{\protect\rule{.1in}{.1in}}

\newcommand{\rot}{\mathop{\rm rot}\nolimits}

\begin{document}
\title{Collective Behavior of Self Propelling Particles with Kinematic Constraints;
The relation between the discrete and the continuous description}
\author{V.I. Ratushnaya}
\affiliation{Colloid and Interface Science group, LIC, Leiden University, P.O. Box 9502,
2300 RA Leiden, The Netherlands}
\author{D. Bedeaux}
\affiliation{Colloid and Interface Science group, LIC, Leiden University, P.O. Box 9502,
2300 RA Leiden, The Netherlands}
\author{V.L. Kulinskii}
\affiliation{Department for Theoretical Physics, Odessa National
University, Dvoryanskaya 2, 65026 Odessa, Ukraine}
\author{A.V. Zvelindovsky}
\affiliation{Department of Physics, Astronomy \& Mathematics, University of Central
Lancashire, Preston PR1 2HE, United Kingdom}

\begin{abstract}
In two papers we proposed a continuum model for the dynamics of systems of
self propelling particles with kinematic constraints on the velocities and
discussed some of its properties. The model aims to be analogous to a discrete
algorithm used in works by T. Vicsek et al. In this paper we derive the
continuous hydrodynamic model from the discrete description. The similarities
and differences between the resulting model and the hydrodynamic model
postulated in our previous papers are discussed. The results clarify the
assumptions used to obtain a continuous description.

\end{abstract}
\maketitle






\section{Introduction}

Systems of self propelling particles (SPP) are widely represented in
the nature. Flocks of birds, schools of fishes, swarms of ants,
groups of bacteria etc. are examples of such systems
\cite{camazine2001,parrishbook1997,parrish1999,parrish2002}. The
observed complex behavior of these systems is far from completely
understood. That is why this phenomenon is of great interest.
Pattern formation in the systems of SPP is caused either by the
presence of external fields (e.g. temperature gradient, food density
gradient, chemotaxis phenomenon etc.) or by kinematic constraints
which are imposed on the motion of the particles. In this second
case the clustering of SPP is driven by internal dynamics of a
nonpotential character. The tendency of the particles to align their
velocities with their neighbors is a crucial element in the
mechanism of the emergence of the coherent motion of the SPP.
Numerous attempts have been made to find a model describing the
collective behavior of self propelling particles. One may
distinguish two main directions of research: the numerical
simulations (discrete description) and the hydrodynamic (continuous)
approaches.

The first numerical model (discrete algorithm) for coherent motion
of SPP was proposed by T. Vicsek et al. \cite{cva/prl/1995}. In
their work a simple kinematic updating rule for the directions of
the velocities of the particles was proposed and numerical evidence
was given for a transition from a disordered state to ordered
collective motion at low noise amplitude and high density values.
They seem to have been the first to realize that flocks fall into a
broad category of nonequilibrium dynamical systems with many degrees
of freedom and noted the analogy between flocking and
ferromagnetism. The model of T. Vicsek has become the ``minimal``
simulation model in the ``modern era`` of the study of flocking.
Extensions of T. Vicsek's model have been proposed considering
particles with varying speeds and different types of noise, by
inclusion of external force fields and/or interparticle attractive
and repulsive forces
\cite{cva/1996/pre,ecolmod2004,gregchate/2003/phd,gregchate/prl/2004}.

Properties of T. Vicsek's model were also investigated from a
mathematical point of view. In \cite{tanner2005} the spontaneous
emergence of ordered motion has been studied in terms of so called
control laws using graph theory. Generalizations of the control laws
were considered in \cite{jad2003,sepulchre2005}. In
\cite{sepulchre2005} it was shown that the organized motion of SPP
with the control laws depending on the relative orientations of the
velocities and relative spacing, can be of two types only: parallel
and circular motion. The stability properties of these discrete
updating rules (including the T.Vicsek's model) and the dynamics
they describe were considered using Lyapunov's theory in
\cite{tanner2005,jad2003,gazi2003,gazi2004}.

The work by T.Vicsek et al. also gave an impulse to the development of
continuous approaches. Here one may distinguish two classes of approaches.

The first class consists of the models which are usually based on
the analogy with the Navier-Stokes equation, where terms describing
the self propelling nature of the system are added. In
\cite{cv/pha/2000} the pressure and viscous terms are incorporated
into the model side by side with the driving force and the friction
caused by the interaction with the environment. The inclusion of
additional terms in \cite{ttr/annph/2005,ramasw/prl/2002} was done
based on symmetry consideration. The attempt to derive such
phenomenological equations from the kinetic equation was made
recently in \cite{gregorie/arxiv/2006}.

The second class contains models which describe the swarming behavior of SPP
by inclusion of attractive and repulsive interactions. The model, based on the
diffusion-advection-reaction equation with nonlocal attractive and repulsive
terms, is suggested in \cite{Mogilner1999} in order to describe the swarming
phenomenon. Their model gives a compact (with sharp edges) aggregation of SPP
with a constant density as a result, which according to the authors is
biologically reasonable.

Another continuous model for the behavior of the living organisms
with nonlocal social interactions is proposed in
\cite{topaz/siam/2004}. There the kinematic rule for the velocity
field contains the density dependent drift and the nonlocal
attraction and repulsion contributions. For a 2-dimensional case of
incompressible fluid with the motion of the particles being normal
to the population gradients, the flow of a constant density with a
compact support is obtained.

In two papers \cite{usepll2005,usphysica} we proposed a hydrodynamic
model, which can be considered to be the continuum analogue of the
discrete dynamic automaton proposed by T. Vicsek et al.
\cite{cva/prl/1995}, which we further will call the CVA model or
algorithm. We constructed our model on the basis of the physical
properties of the discrete CVA model, namely the conservation of the
number of particles and the kinetic energy. The discrete
configuration updating rule used by T. Vicsek et al. changes only
the direction of the particle velocities but keeps their absolute
value constant. In their algorithm the number of particles is
constant as well.

In this article we obtain the continuous description by coarse graining the
discrete CVA algorithm. In this respect our present paper is meant to be a
link between two existing groups of approaches: discrete and continuous. The
importance of this analysis is that it clarifies which of the continuous
models we proposed is closest to be the continuum analog of the CVA model.

In Section 2 we will start with a rule for the velocities formulated by T.
Vicsek et al. and obtain a discrete equation of motion for each particle. We
introduce angular velocities associated with the rate of change of the
direction of the linear velocity of the particles. These angular velocities
contain the information about the nonpotential interactions between a given
particle and its local surrounding. We derive an expression for the angular
velocities in the continuous time description. We show that to a first order
in the velocity difference between the steps the angular velocity for particle
$i$\ depends on the average velocity in the neighborhood of the $i$th particle
and its rate of change.

In Section 3 we obtain the continuous description, with a conserved kinetic
energy and number of particles, using a coarse-graining procedure. We obtain
the angular velocity field that follows from the CVA 2-dimensional model and
compare it with the angular velocity fields we proposed in our first paper. It
turns out that there are similarities and differences. Both the continuous
description that follows from the CVA model and our continuous model are and
have been shown to give stationary linear and vortical flow fields. The
description of such flow fields is one of the aims of the model. A discussion
is given and concluding remarks are made in the last section.

\section{Continuum time limit}

In this section we derive the equation of motion in continuous time
from the CVA algorithm. In their work \cite{cva/prl/1995} the
collective behavior of self propelling particles with respect to a
change of the density and the strength of the noise was
investigated. In our analysis the noise will not be considered. We
focus on the systematic contribution. In our first paper we
discussed how noise can be added in our approach.

The ordered motion of self propelling particles in
\cite{cva/prl/1995} is described by the CVA rule, according to which
at each discrete time step (labeled by $n$) the $i$th particle
adjusts the direction of its velocity $\mathbf{v}_{i}\left(
n\right)  $ to the direction of the average velocity
$\mathbf{u}_{i}\left(  n\right)  $ in its neighborhood. The average
is calculated over a region with a radius $R$ around a given
particle. Using this radius we will call particle densities small
compared to $R^{-d}$, where $d$ is the dimensionality, small. When
the density is larger we call it large. The CVA rule implies that
\begin{equation}
\mathbf{v}_{i}\left(  n+1\right)  \times\mathbf{u}_{i}\left(  n\right)
=0,\quad\forall\,i,n\,, \label{cva}%
\end{equation}
where the absolute value of the velocity of each particle is assumed to be
constant , i.e.
\begin{equation}
\mid\mathbf{v}_{i}\left(  n+1\right)  \mid=\mid\mathbf{v}_{i}\left(  n\right)
\mid=\mathrm{v}\mathit{_{i}\,.} \label{2}%
\end{equation}
Together with Eq.~\eqref{cva} it follows that
\begin{equation}
\mathbf{v}_{i}\left(  n+1\right)
=\mathbf{v}_{i}\,\mathbf{u}_{i}\left( n\right) ,
\quad\text{where}\quad\mid\mathbf{u}_{i}\left( n \right) \mid =1\,.%
\end{equation}
Using the fact that $\mathbf{v}_{i}\left(  n+1\right)  -\mathbf{v}_{i}\left(
n\right)  $ is perpendicular to $\mathbf{v}_{i}\left(  n+1\right)
+\mathbf{v}_{i}\left(  n\right)  $, given the validity of Eq.~\eqref{2}, it
can be shown that
\begin{equation}
\mathbf{v}_{i}\left(  n+1\right)  -\mathbf{v}_{i}\left(  n\right)  =\left[
\widehat{\mathbf{v}}_{i}\left(  n\right)  \times\widehat{\mathbf{v}}%
_{i}\left(  n+1\right)  \right]  \times\left[  \frac{\mathbf{v}_{i}\left(
n+1\right)  +\mathbf{v}_{i}\left(  n\right)  }{1+\widehat{\mathbf{v}}%
_{i}\left(  n\right)  \cdot\widehat{\mathbf{v}}_{i}\left(  n+1\right)
}\right]  , \label{dif}%
\end{equation}
where $\widehat{\mathbf{v}}_{i}\left(  n\right)  \equiv\mathbf{v}_{i}\left(
n\right)  /\mathit{{v}_{i}}$ is a unit vector in the direction of the velocity
$\mathbf{v}_{i}\left(  n\right)  $.

It is important to realize that there is a difference between low density
regions and high density regions. In high density regions the velocity of the
particles is updated at every step. In the low density regions the average of
the velocity of particles around and including particle $i$ is equal to the
velocity of particle $i$. It follows that $\mathbf{u}_{i}\left(  n\right)
=\mathbf{v}_{i}\left(  n\right)  /\mathrm{v}\mathit{_{i}}$. As a consequence
$\mathbf{v}_{i}\left(  n+1\right)  =\mathbf{v}_{i}\left(  n\right)  $.
Substitution in Eq.~\eqref{dif} gives the equality zero equal to zero. The
important conclusion is that in the low density regions the particles do not
change their velocity. We will come back to this point when this is relevant.

In order to obtain a continuous description as a function of time, we assume
the steps to be small so that
\begin{equation}
\mid\widehat{\mathbf{v}}_{i}\left(  n+1\right)  -\widehat{\mathbf{v}}%
_{i}\left(  n\right)  \mid\ll1\,. \label{cond}%
\end{equation}
One may then write Eq.~\eqref{dif} to first order in the velocity difference
as
\begin{equation}
\mathbf{v}_{i}\left(  n+1\right)  -\mathbf{v}_{i}\left(  n\right)  =\left[
\widehat{\mathbf{v}}_{i}\left(  n\right)  \times\widehat{\mathbf{v}}%
_{i}\left(  n+1\right)  \right]  \times\mathbf{v}_{i}\left(  n\right)
=\left[  \widehat{\mathbf{v}}_{i}\left(  n\right)  \times\mathbf{u}_{i}\left(
n\right)  \right]  \times\mathbf{v}_{i}\left(  n\right)  \,. \label{veldif}%
\end{equation}
As we are interested in the rate of change of the velocity we divide this
equation by the time step duration $\tau$. This gives
\begin{equation}
\frac{\mathbf{v}_{i}\left(  n+1\right)  -\mathbf{v}_{i}\left(  n\right)
}{\tau}=\left[  \frac{\widehat{\mathbf{v}}_{i}\left(  n\right)  \times
\mathbf{u}_{i}\left(  n\right)  }{\tau}\right]  \times\mathbf{v}_{i}\left(
n\right)  =\boldsymbol{\omega}_{\mathbf{v}_{i}}\times\mathbf{v}_{i}\,,
\label{der}%
\end{equation}
where
\begin{equation}
\boldsymbol{\omega}_{\mathbf{v}_{i}}=\frac{1}{\tau}\,\widehat{\mathbf{v}}%
_{i}\times\mathbf{u}_{i}%
\end{equation}
is an angular velocity associated with the velocity vector $\mathbf{v}_{i}$.

In view of Eq.~\eqref{cond} the left hand side of Eq.~\eqref{der} gives the
continuous time derivative. In other words we may introduce the following
definition:
\begin{equation}
\dot{\mathbf{v}}_{i}\left(  n\right)  \longleftrightarrow\frac{\mathbf{v}%
_{i}\left(  n+1\right)  -\mathbf{v}_{i}\left(  n\right)  }{\tau}\,.
\end{equation}
Using that $\mathbf{u}_{i}\left(  n\right)  =\widehat{\mathbf{v}}_{i}\left(
n+1\right)  $ it follows from Eq.~\eqref{der} that
\begin{equation}
\mathbf{u}_{i}\left(  n+1\right)  -\mathbf{u}_{i}\left(  n\right)
=\tau\,\boldsymbol{\omega}_{\mathbf{u}_{i}}\left(  n\right)  \times
\mathbf{u}_{i}\left(  n\right)  \,,
\end{equation}
where the angular velocity $\boldsymbol{\omega}_{\mathbf{u}_{i}}\left(
n\right)  $ corresponding to the average velocity $\mathbf{u}_{i}\left(
n\right)  $ is defined as
\begin{equation}
\boldsymbol{\omega}_{\mathbf{u}_{i}}\left(  n\right)  =\boldsymbol{\omega
}_{\mathbf{v}_{i}}\left(  n+1\right)  .
\end{equation}
It can be shown that
\begin{equation}
\boldsymbol{\omega}_{\mathbf{u}_{i}}\left(  n\right)  =\frac{1}{\tau}\left[
\widehat{\mathbf{v}}_{i}\left(  n+1\right)  \times\mathbf{u}_{i}\left(
n+1\right)  \right]  =\frac{1}{\tau}\left[  \mathbf{u}_{i}\left(  n\right)
\times\mathbf{u}_{i}\left(  n+1\right)  \right]  =\mathbf{u}_{i}\left(
n\right)  \times\dot{\mathbf{u}}_{i}\left(  n\right)  \label{omudiscr}%
\end{equation}
where
\begin{equation}
\dot{\mathbf{u}}_{i}\left(  n\right)  =\frac{\mathbf{u}_{i}\left(  n+1\right)
-\mathbf{u}_{i}\left(  n\right)  }{\tau}\,.
\end{equation}
Furthermore one may show that
\begin{equation}
\boldsymbol{\omega}_{\mathbf{u}_{i}}\left(  n\right)  -\boldsymbol{\omega
}_{\mathbf{v}_{i}}\left(  n\right)  =\tau\,\widehat{\mathbf{v}}_{i}\left(
n+1\right)  \times\ddot{\widehat{\mathbf{v}}}_{i}\left(  n\right)
=\tau\,\mathbf{u}_{i}\left(  n\right)  \times\ddot{\widehat{\mathbf{v}}}%
_{i}\left(  n\right)  \,, \label{omdiff}%
\end{equation}
where the second order derivative is defined by
\begin{equation}
\ddot{\mathbf{v}}_{i}\left(  n\right)  =\frac{1}{\tau^{2}}\Big[\mathbf{v}%
_{i}\left(  n+2\right)  -2\mathbf{v}_{i}\left(  n+1\right)  +\mathbf{v}%
_{i}\left(  n\right)  \Big].
\end{equation}
Combining Eqs.~\eqref{omudiscr} and \eqref{omdiff} results in
\begin{equation}
\boldsymbol{\omega}_{\mathbf{v}_{i}}\left(  n\right)  =\boldsymbol{\omega
}_{\mathbf{u}_{i}}\left(  n\right)  -\tau\,\mathbf{u}_{i}\left(  n\right)
\times\ddot{\widehat{\mathbf{v}}}_{i}\left(  n\right)  =\mathbf{u}_{i}\left(
n\right)  \times\dot{\mathbf{u}}_{i}\left(  n\right)  -\tau\,\mathbf{u}%
_{i}\left(  n\right)  \times\ddot{\widehat{\mathbf{v}}}_{i}\left(  n\right)
\,,
\end{equation}
which to first order in the velocity difference implies that for the angular
velocity associated with the particle velocity $\mathbf{v}_{i}$ we obtain the
following expression:
\begin{equation}
\boldsymbol{\omega}_{\mathbf{v}_{i}}\left(  n\right)  =\boldsymbol{\omega
}_{\mathbf{u}_{i}}\left(  n\right)  -\tau\,\mathbf{u}_{i}\left(  n\right)
\times\ddot{\widehat{\mathbf{u}}}_{i}\left(  n\right)  =\mathbf{u}_{i}\left(
n\right)  \times\dot{\mathbf{u}}_{i}\left(  n\right)  \,. \label{omega}%
\end{equation}
The second equality follows from the fact that the second derivative
$\ddot{\mathbf{u}}_{i}\left(  n\right)  $ is parallel to $\mathbf{u}%
_{i}\left(  n\right)  $ to first order in the difference.

Replacing $n$ by the time $t$ the resulting equation of motion becomes:
\begin{equation}
\frac{d\mathbf{v}_{i}(t)}{dt}=\boldsymbol{\omega}_{\mathbf{u}_{i}}%
(t)\times\mathbf{v}_{i}(t)=\left[  \mathbf{u}_{i}(t)\times\dot{\mathbf{u}}%
_{i}(t)\right]  \times\mathbf{v}_{i}(t). \label{eqmot}%
\end{equation}
This equation is continuous in time and is derived from the discrete CVA rule
using Eq.~\eqref{cond}. In order to obtain equations for the velocity and the
density fields, which are continuous in space, we will coarse-grain
Eqs.~\eqref{omega} and \eqref{eqmot} in the next section. We note that both
$d\mathbf{v}_{i}\left(  t\right)  /dt$ and $\dot{\mathbf{u}}_{i} \left(
t\right)  $ are zero in the low density regions.

\section{Continuous transport equations}

\label{sectrprt} In this section we introduce the averaging procedure for the
discrete model. We derive continuous equations for the velocity and the
density fields by averaging the discrete equations.

In our continuum model the number density, $n(\mathbf{r},t),$ satisfies the
continuity equation,
\begin{equation}
\frac{\partial n(\mathbf{r},t)}{\partial t}+\mathop{\rm div}\nolimits\left(
n(\mathbf{r},t)\,\mathbf{v}(\mathbf{r},t)\right)  =0\,. \label{1.2}%
\end{equation}
The dynamics of the velocity field is such that the kinetic energy density is
conserved, $d|\mathbf{v}(\mathbf{r},t)|^{2}/dt=0$. The time derivative of the
velocity field is therefore given by
\begin{equation}
\frac{d}{dt}\mathbf{v}\left(  \mathbf{r},t\right)  =\frac{\partial}{\partial
t}\mathbf{v}\left(  \mathbf{r},t\right)  +\mathbf{v}\left(  \mathbf{r}%
,t\right)  \cdot\mathrm{grad}\mathbf{v}\left(  \mathbf{r},t\right)
=\boldsymbol{\omega}\left(  \mathbf{r},t\right)  \times\mathbf{v}\left(
\mathbf{r},t\right)  . \label{1.3}%
\end{equation}
where $\boldsymbol{\omega}\left(  \mathbf{r},t\right)  $ is some angular
velocity field. We will discuss how to obtain this angular velocity field in
terms of the velocity and density fields below.

In the CVA algorithm the direction of the average velocity in the neighborhood
of particle $i$ is given for the continuous time description by
\begin{equation}
\mathbf{u}_{i}\left(  t\right)  =\sum\limits_{j}H\left(  \mathbf{r}%
_{ij}\left(  t\right)  \right)  \,\mathbf{v}_{j}\left(  t\right)
\Big|\sum\limits_{j}H\left(  \mathbf{r}_{ij}\left(  t\right)  \right)
\,\mathbf{v}_{j}\left(  t\right)  \Big|^{-1}\,, \label{u}%
\end{equation}
where $r_{ij}=|\mathbf{r}_{i}-\mathbf{r}_{j}|$. The dynamics of individual
particles therefore reduces the difference between the direction of its
velocity and that of the average velocity of the surrounding particles.
$H\left(  \mathbf{r}\right)  $ is an averaging kernel, which we assume to be
normalized,
\begin{equation}
\int H\left(  \mathbf{r}\right)  d\mathbf{r}=1\,.
\end{equation}
It has the characteristic averaging scale $R$, beyond which the
kernel goes to zero \cite{cva/prl/1995,cv/pha/2000}. Usually one
uses for $H$ a normalized Heaviside step function.

In order to obtain a continuous description we define the average particle
density (per unit of volume) and velocity fields by
\begin{align}
n\left(  \mathbf{r},t\right)   &  =\sum_{j}H\left(  \mathbf{r}-\mathbf{r}%
_{j}(t)\right)  ,\nonumber\label{sys}\\
n\left(  \mathbf{r},t\right)  \mathbf{v}\left(  \mathbf{r},t\right)   &
=\sum_{j}H\left(  \mathbf{r}-\mathbf{r}_{j}(t)\right)  \mathbf{v}_{j}(t)\,.
\end{align}
Using Eq.\eqref{sys} in Eq.\eqref{u} for $\mathbf{r}=\mathbf{r}_{i}$, it
follows that $\mathbf{u}_{i}\left(  t\right)  =\widehat{\mathbf{v}}\left(
\mathbf{r}_{i}\left(  t\right)  ,t\right)  $, where $\widehat{\mathbf{v}%
}\left(  \mathbf{r}_{i}\left(  t\right)  ,t\right)  =\mathbf{v}\left(
\mathbf{r}_{i}\left(  t\right)  ,t\right)  /|\mathbf{v}\left(  \mathbf{r}%
_{i}\left(  t\right)  ,t\right)  |$. Eq.~\eqref{eqmot} can therefore be
written as
\begin{equation}
\frac{d\mathbf{v}_{i}(t)}{dt}=\left[  \widehat{\mathbf{v}}\left(
\mathbf{r}_{i}(t),t\right)  \times\frac{d\widehat{\mathbf{v}}\left(
\mathbf{r}_{i}(t),t\right)  }{dt}\right]  \times\mathbf{v}_{i}\left(
t\right)  .
\end{equation}
By evaluating the time derivative between the square brackets on the right
hand side one obtains
\begin{align}
\frac{d\mathbf{v}_{i}(t)}{dt}  &  =\left[  \widehat{\mathbf{v}}\left(
\mathbf{r},t\right)  \times\left(  \mathbf{v}_{i}\left(  t\right)  \cdot
\nabla\widehat{\mathbf{v}}\left(  \mathbf{r},t\right)  \right)  \right]
_{\mathbf{r}=\mathbf{r}_{i}(t)}\times\mathbf{v}_{i}(t)+\nonumber\label{grad}\\
&  \left[  \widehat{\mathbf{v}}\left(  \mathbf{r},t\right)  \times
\frac{\partial\widehat{\mathbf{v}}\left(  \mathbf{r},t\right)  }{\partial
t}\right]  _{\mathbf{r}=\mathbf{r}_{i}(t)}\times\mathbf{v}_{i}(t)=\nonumber\\
& \nonumber\\
&  \left[  \widehat{\mathbf{v}}\left(  \mathbf{r},t\right)  \times\left[
\left(  \mathbf{v}_{i}\left(  t\right)  -\mathbf{v}(\mathbf{r},t)\right)
\cdot\nabla\widehat{\mathbf{v}}\left(  \mathbf{r},t\right)  \right)  \right]
_{\mathbf{r}=\mathbf{r}_{i}(t)}\times\mathbf{v}_{i}(t)+\nonumber\\
&  \left[  \widehat{\mathbf{v}}\left(  \mathbf{r},t\right)  \times
\frac{d\widehat{\mathbf{v}}\left(  \mathbf{r},t\right)  }{dt}\right]
_{\mathbf{r}=\mathbf{r}_{i}(t)}\times\mathbf{v}_{i}(t)\,.
\end{align}
In view of the fact that the gradient of the direction of the average velocity
is of the first order and that the velocity difference is also of the first
order, the first contribution on the right hand side is of the second order
and can be neglected. Eq.~\eqref{grad} therefore reduces to
\begin{equation}
\frac{d\mathbf{v}_{i}(t)}{dt}=\left[  \widehat{\mathbf{v}}\left(
\mathbf{r},t\right)  \times\frac{d\widehat{\mathbf{v}}\left(  \mathbf{r}%
,t\right)  }{dt}\right]  _{\mathbf{r}=\mathbf{r}_{i}(t)}\times\mathbf{v}%
_{i}(t)\,. \label{vi}%
\end{equation}
Note that both $d\mathbf{v}_{i}\left( t\right) /dt$ and $d\widehat{\mathbf{v}
}\left( \mathbf{r}_{i}\left( t\right) ,t\right) /dt$ are zero in the low
density regions. When we now average this equation we can use the fact that
the expression between the square brackets only depends on the coarse grained
functions and therefore varies slowly over the range of the averaging
function.\\*Averaging the left hand side of Eq.~\eqref{vi} and using the
continuity equation we obtain
\begin{align}
&  \sum_{i}\frac{d\mathbf{v}_{i}\left(  t\right)  }{dt}H\left(  \mathbf{r}%
-\mathbf{r}_{i}\left(  t\right)  \right)  =\nonumber\\
&  \frac{\partial}{\partial t}\sum\limits_{i}\mathbf{v}_{i}\left(  t\right)
H\left(  \mathbf{r}-\mathbf{r}_{i}\left(  t\right)  \right)  -\sum
\limits_{i}\mathbf{v}_{i}\left(  t\right)  \frac{\partial}{\partial
\mathbf{r}_{i}}\cdot\mathbf{v}_{i}\left(  t\right)  H\left(  \mathbf{r}%
-\mathbf{r}_{i}\left(  t\right)  \right)  =\nonumber\\
&  \frac{\partial\left(  n\left(  \mathbf{r},t\right)  \mathbf{v}\left(
\mathbf{r},t\right)  \right)  }{\partial t}+\frac{\partial}{\partial
\mathbf{r}}\cdot\sum\limits_{i}\mathbf{v}_{i}\left(  t\right)  \,\mathbf{v}%
_{i}\left(  t\right)  \,H\left(  \mathbf{r}-\mathbf{r}_{i}\left(  t\right)
\right)  =\nonumber\\
&  n\left(  \mathbf{r},t\right)  \,\frac{d\mathbf{v}\left(  \mathbf{r}%
,t\right)  }{dt}-\nabla\cdot\left[  n\left(  \mathbf{r},t\right)
\,\mathbf{v}\left(  \mathbf{r},t\right)  \otimes\mathbf{v}\left(
\mathbf{r},t\right)  \right]  +\nonumber\\
&  \ \nabla\cdot\sum\limits_{i}\mathbf{v}_{i}\left(  t\right)  \mathbf{v}%
_{i}\left(  t\right)  \,H\left(  \mathbf{r}-\mathbf{r}_{i}\left(  t\right)
\right)  =n\left(  \mathbf{r},t\right)  \,\frac{d\mathbf{v}\left(
\mathbf{r},t\right)  }{dt}+\nonumber\\
&  \nabla\cdot\sum\limits_{i}\left(  \mathbf{v}\left(  \mathbf{r},t\right)
-\mathbf{v}_{i}\left(  t\right)  \right)  \left(  \mathbf{v}\left(
\mathbf{r},t\right)  -\mathbf{v}_{i}\left(  t\right)  \right)  \,H\left(
\mathbf{r}-\mathbf{r}_{i}\left(  t\right)  \right)  =\nonumber\\
&  n\left(  \mathbf{r},t\right)  \,\frac{d\mathbf{v}\left(  \mathbf{r}%
,t\right)  }{dt}\,,
\end{align}
where we neglected the term of the second order in the velocity difference.
This term would give a small contribution to the pressure tensor. It is
therefore also a contribution which is in the formulation of the problem
assumed to be cancelled by the self propelling force.

In the right hand side of Eq.~\eqref{vi} we have
\begin{align}
&  \sum\limits_{i}\left[  \widehat{\mathbf{v}}\left(  \mathbf{r},t\right)
\times\frac{d\widehat{\mathbf{v}}\left(  \mathbf{r},t\right)  }{dt}\right]
_{\mathbf{r}=\mathbf{r}_{i}(t)}\times\mathbf{v}_{i}\left(  t\right)
\,H\left(  \mathbf{r}-\mathbf{r}_{i}\left(  t\right)  \right)  =\nonumber\\
&  \left[  \widehat{\mathbf{v}}\left(  \mathbf{r},t\right)  \times
\frac{d\widehat{\mathbf{v}}\left(  \mathbf{r},t\right)  }{dt}\right]  \times
n\left(  \mathbf{r},t\right)  \mathbf{v}\left(  \mathbf{r},t\right)  .
\end{align}
This implies that the averaged equation of motion can be written as
\begin{equation}
\frac{d\mathbf{v}\left(  \mathbf{r},t\right)  }{dt}=\left[  \widehat
{\mathbf{v}}\left(  \mathbf{r},t\right)  \times\frac{d\widehat{\mathbf{v}%
}\left(  \mathbf{r},t\right)  }{dt}\right]  \times\mathbf{v}\left(
\mathbf{r},t\right)  \,. \label{eomaver}%
\end{equation}
This gives
\begin{equation}
\frac{d\mathbf{v}\left(  \mathbf{r},t\right)  }{dt}=\boldsymbol{\omega}\left(
\mathbf{r},t\right)  \times\mathbf{v}\left(  \mathbf{r},t\right)  \,,
\label{eqmotion}%
\end{equation}
where to first order in the velocity difference
\begin{equation}
\boldsymbol{\omega}\left(  \mathbf{r},t\right)  =\widehat{\mathbf{v}}\left(
\mathbf{r},t\right)  \times\frac{d\widehat{\mathbf{v}}\left(  \mathbf{r}%
,t\right)  }{dt}=\frac{1}{\mathrm{v}^{2}\left(  \mathbf{r},t\right)  }\left[
\mathbf{v}\left(  \mathbf{r},t\right)  \times\frac{d\mathbf{v}\left(
\mathbf{r},t\right)  }{dt}\right]  \,. \label{omega2}%
\end{equation}
where $\mathrm{v}\left(  \mathbf{r},t\right)  \equiv|\mathbf{v}\left(
\mathbf{r},t\right)  |$.

Here we restrict our discussion by considering the 2-dimensional case in order
to make a comparison with the results obtained in our previous papers. By
evaluating the time derivative in Eq.~\eqref{omega2} we may rewrite this
expression as follows
\begin{align}
\boldsymbol{\omega}\left(  \mathbf{r},t\right)   &  =\frac{1}{\mathrm{v}%
^{2}\left(  \mathbf{r},t\right)  }\left[  \mathbf{v}\left(  \mathbf{r}%
,t\right)  \times\left(  \frac{\partial\mathbf{v}\left(  \mathbf{r},t\right)
}{\partial t}+\left(  \mathbf{v}\left(  \mathbf{r},t\right)  \cdot
\nabla\right)  \mathbf{v}\left(  \mathbf{r},t\right)  \right)  \right]
=\nonumber\\
& \nonumber\\
&  \frac{1}{\mathrm{v}^{2}\left(  \mathbf{r},t\right)  }\left[  \mathbf{v}%
\left(  \mathbf{r},t\right)  \times\left(  \frac{\partial\mathbf{v}\left(
\mathbf{r},t\right)  }{\partial t}+\nabla\frac{\mathrm{v}^{2}\left(
\mathbf{r},t\right)  }{2}\right)  \right]  -\nonumber\\
& \nonumber\\
&  \frac{1}{\mathrm{v}^{2}\left(  \mathbf{r},t\right)
}\,\mathbf{v}\left( \mathbf{r},t\right)  \times\left[
\mathbf{v}\left(  \mathbf{r},t\right)
\times\rot\nolimits\mathbf{v}\left(  \mathbf{r},t\right)  \right]
=\nonumber\\
& \nonumber\\
&  \frac{\mathbf{v}\left(  \mathbf{r},t\right)  }{\mathrm{v}^{2}\left(
\mathbf{r},t\right)  }\times\frac{\partial\mathbf{v}\left(  \mathbf{r}%
,t\right)  }{\partial t}+\frac{\mathbf{v}\left(  \mathbf{r},t\right)
}{\mathrm{v}^{2}\left(  \mathbf{r},t\right)  }\times\nabla\frac{\mathrm{v}%
^{2}\left(  \mathbf{r},t\right)  }{2}+\rot\nolimits\mathbf{v}%
\left(  \mathbf{r},t\right)  \,.\label{omegaCVA}%
\end{align}
This is the angular velocity field obtained from the discrete algorithm used
by T. Vicsek et al.

One may see that the continuous equation of motion, Eq.~\eqref{eqmotion}, with
the angular velocity derived from the CVA rule, Eq.~\eqref{omegaCVA}, can be
written as follows:
\begin{equation}
\frac{d\mathbf{v}}{dt}=\left(
\mathbf{1}-\widehat{\mathbf{v}}\widehat {\mathbf{v}}\right)
\cdot\frac{\partial\mathbf{v}}{\partial t}+\left(
\mathbf{1}-\widehat{\mathbf{v}}\widehat{\mathbf{v}}\right)
\cdot\nabla \frac{\mathrm{v}^{2}}{2}+\left(
\rot\nolimits\mathbf{v}\right)
\times\mathbf{v}. \label{eomaver2}%
\end{equation}
where $\mathbf{1}$\ is the unit tensor. All three terms on the right hand side
contribute to the co-moving derivative of the velocity which is orthogonal to
the velocity field.

In the low density regions one obtains, as has been pointed out a number of
times, $d\mathbf{v}/dt=0$. As this is not so clearly visible in
Eq.\eqref{eomaver2} it is appropriate to replace this equation by
\begin{equation}
\frac{d\mathbf{v}}{dt}=\left[  \left(  \mathbf{1}-\widehat{\mathbf{v}}%
\widehat{\mathbf{v}}\right)  \cdot\frac{\partial\mathbf{v}}{\partial
t}+\left(
\mathbf{1}-\widehat{\mathbf{v}}\widehat{\mathbf{v}}\right)
\cdot\nabla\frac{\mathrm{v}^{2}}{2}+\left(
\rot\nolimits\mathbf{v}\right)  \times\mathbf{v}\right]  \text{
}f(n)\,, \label{eomaver3}%
\end{equation}
where $f(n)$ is the density dependent factor, which arises due to
coarse-graining procedure. In low density limit it is natural that
$f(n)\to 0$ as $n\to 0$. A more thorough analysis of this is beyond
our present aim, however.

Before comparing this expression to the one we used in Refs.~
\cite{usepll2005,usphysica} we first verify that stationary linear
and the vortical solutions are solutions of Eq.~\eqref{eomaver2}. In
view of their stationarity the first contribution in
Eq.~\eqref{eomaver2} is equal to zero. For a linear flow
$\mathbf{v}=\mathrm{v}_{0}\,\mathbf{e}_{x}$ the other two terms and
the left hand side of Eq.~\eqref{eomaver2} are also zero. Stationary
linear flow is therefore a solution. In case of stationary vortical
flow,
$\mathbf{v}=\mathrm{v}_{\varphi}(r)\,\mathbf{e}_{\varphi}\left(
\varphi\right) $, the $\mathbf{v}\cdot\nabla\mathbf{v}$ term on the
left hand side of Eq.~\eqref{eomaver2} cancels the terms due to the
second and the third term. The continuity equation, Eq.~\eqref{1.2},
is satisfied for each density distribution which varies only in
directions normal to the flow direction. It follows that the
continuous analog of the CVA model has stationary linear and
vortical solutions.

In our first paper \cite{usepll2005} we used an angular velocity
field which was a linear combination of $n\left(
\mathbf{r},t\right) \rot\nolimits\mathbf{v}\left(
\mathbf{r},t\right)  $ and $\nabla n\left(  \mathbf{r},t\right)
\times\mathbf{v}\left(  \mathbf{r},t\right)  $. The resulting
equation of motion becomes
\begin{equation}
\frac{d\mathbf{v}}{dt}=s_{1}n\left(  \rot\nolimits\mathbf{v}%
\right)  \times\mathbf{v}+s_{2}\left(  \nabla n\times\mathbf{v}\right)
\times\mathbf{v.} \label{us}%
\end{equation}
The first term is analogous to the third term on the right hand side
of Eq.~\eqref{eomaver2}. Similar to the CVA model this choice leads
to stationary linear and vortical solutions. The linear dependence
of our choice on the density leads to a dependence of the stationary
velocity field on the density distribution. We refer to
\cite{usepll2005,usphysica} for a detailed discussion of these
solutions. For a small density the right hand side of Eq.~\eqref{us}
makes $d\mathbf{v/}dt$ negligible. This is similar to the behavior
in Eq.~\eqref{eomaver3}.

When one modifies the updating rule in the CVA model, as done in
Refs.~\cite{cva/1996/pre, ecolmod2004,gregchate/prl/2004}, this
leads to a modification of the $\boldsymbol{\omega}$ given in
Eq.~\eqref{omegaCVA}. Similarly, the choice of $\boldsymbol{\omega}$
we used in \cite{usepll2005, usphysica} can be modified to include
such contributions. The freedom in the choice of
$\boldsymbol{\omega}$\ in the continuous version of the CVA model is
one of its strength.

\section{Conclusions}

\label{secncls} In our first paper \cite{usepll2005} we constructed
a continuous self propelling particle model with particle number and
kinetic energy conservation. In this paper we addressed the problem
to derive the continuous description from the discrete model
proposed by T. Vicsek et al. \cite{cva/prl/1995}. We were able to
derive expressions for the angular velocity field used in the
continuous model from the updating procedures used in their model.
By coarse graining the discrete equations in the original model we
obtained the angular velocity field used to give the co-moving time
derivative of the velocity field in the continuous description.
Modification of the updating rules in this model, as done in
Refs.~\cite{cva/1996/pre,ecolmod2004,gregchate/2003/phd,gregchate/prl/2004},
results in modifications of the resulting angular velocity field.
The angular velocity field used in our work
\cite{usepll2005,usphysica} is one of such choices. One of the
contributions in the continuous version of the CVA model is very
similar to one of the contributions which we have postulated in our
hydrodynamic model \cite{usepll2005}. Both the continuous CVA model
and our model lead to the linear and vortical flows of the self
propelling particles observed in nature and obtained in simulations
and continuum approximations. This shows that they are appropriate
for the description of flocking behavior, which is one of the aims
of the model. An
interesting alternative choice of $\boldsymbol{\omega}\left(  \mathbf{r}%
,t\right)  $\ in the continuous description is for instance $\nabla
c_{A}\times\mathbf{v}$ where $c_{A}$ is the concentration of an
attractant. As was shown in A. Czir\'{o}k et al. \cite{cva/1996/pre}
this choice can be used to describe $\mathit{rotor chemotaxis}$.
Note that the term $\nabla n\left( \mathbf{r},t\right)
\times\mathbf{v}\left(  \mathbf{r},t\right)  $, which we considered
in our continuum model, is similar to the one considered by A.
Czir\'{o}k et al. when the concentration field is proportional to
the concentration of the attractant.

Our analysis shows that one may coarse grain the discrete updating rules and
obtain the corresponding continuous description. This makes a direct
comparison between discrete and continuous descriptions possible. For our own
work it was found that our continuous description was similar to the
continuous version of the original CVA model but not identical. The analysis
in this paper makes it possible to extend our work on the continuous
description such that it is either closer or more different from the original
CVA model.\bigskip

\textbf{Acknowledgments}\newline Vladimir Kulinskii thanks the Nederlandse
Organisatie voor Wetenschappelijk Onderzoek (NWO) for a grant, which enabled
him to visit Dick Bedeaux's group at Leiden University.


\end{document}